\documentclass[pmlr,twocolumn,10pt]{jmlr}

\usepackage{longtable}

\usepackage{booktabs}
\usepackage[load-configurations=version-1]{siunitx} 

\theorembodyfont{\upshape}
\theoremheaderfont{\scshape}
\theorempostheader{:}
\theoremsep{\newline}

\jmlrvolume{81}
\jmlryear{2018}
\jmlrworkshop{Conference on Fairness, Accountability, and Transparency}

\title[Fairness in Machine Learning: Lessons from Political Philosophy]{Fairness in Machine Learning: Lessons from Political Philosophy}

\begin{document}

  \author{\Name{Reuben Binns} \Email{reuben.binns@cs.ox.ac.uk}\\
   \addr Department of Computer Science, University of Oxford, United Kingdom}



\editors{Sorelle A. Friedler and Christo Wilson}

\maketitle

\begin{abstract}
What does it mean for a machine learning model to be `fair', in terms which can be operationalised? Should fairness consist of ensuring everyone has an equal probability of obtaining some benefit, or should we aim instead to minimise the harms to the least advantaged? Can the relevant ideal be determined by reference to some alternative state of affairs in which a particular social pattern of discrimination does not exist? Various definitions proposed in recent literature make different assumptions about what terms like discrimination and fairness mean and how they can be defined in mathematical terms.
Questions of discrimination, egalitarianism and justice are of significant interest to moral and political philosophers, who have expended significant efforts in formalising and defending these central concepts. It is therefore unsurprising that attempts to formalise `fairness' in machine learning contain echoes of these old philosophical debates. This paper draws on existing work in moral and political philosophy in order to elucidate emerging debates about fair machine learning.
\end{abstract}

\section{Introduction}
\label{sec:intro}

Machine learning allows us to predict and classify phenomena, by training models using labeled data from the real world. When consequential decisions are made about individuals on the basis of the outputs of such models, concerns about discrimination and fairness inevitably arise. What if the model's outputs result in decisions that are systematically biased against people with certain protected characteristics like race, gender or religion? If there are underlying patterns of discrimination in the real world, such biases will likely be picked up in the learning process. This could result in certain groups being unfairly denied loans, insurance, or employment opportunities. Cognisant of this problem, a burgeoning research paradigm of `discrimination-aware data mining' and `fair machine learning' has emerged, which attempts to detect and mitigate unfairness ~\cite{hajian2013methodology,kamiran2013techniques,barocas2016big}.

One question which immediately arises in such an endeavour is the need for formalisation. What does it mean for a machine learning model to be `fair' or `non-discriminatory', in terms which can be operationalised? Various measures have been proposed. A common theme is comparing differences in treatment between protected and non-protected groups, but there are multiple ways to measure such differences. The most basic would be `disparate impact' or `statistical / demographic parity', which consider the overall percentage of positive/ negative classification rates between groups. However, this is blunt, since it fails to account for discrimination which is explainable in terms of legitimate grounds ~\cite{dwork2012fairness}. For instance, attempting to enforce equal impact between men and women in recidivism prediction systems, if men have higher re-offending rates, could result in women remaining in prison longer despite being less likely to re-offend.

A range of more nuanced measures have been proposed, including; `accuracy equity', which considers the overall accuracy of a predictive model for each group ~\cite{angwin2016machine}; `conditional accuracy equity', which considers the accuracy of a predictive model for each group, conditional on their predicted class ~\cite{dieterich2016compas}; `equality of opportunity', which considers whether each group is equally likely to be predicted a desirable outcome given the actual base rates for that group ~\cite{hardt2016equality}; and `disparate mistreatment', a corollary which considers differences in false positive rates between groups ~\cite{zafar2017fairness}. Another definition involves imagining counterfactual scenarios wherein members of protected groups are instead members of the non-protected group ~\cite{kusner2017counterfactual}. To the extent that outcomes would differ, the system is unfair; i.e. a woman classified by a fair system should get the same classification she would have done in a counterfactual scenario in which she had been born a man.

Ideally, a system might be expected to meet all of these intuitively plausible measures of fairness. But, somewhat problematically, certain measures turn out to be mathematically impossible to satisfy simultaneously except in rare and contrived circumstances ~\cite{kleinberg2016inherent}, and therefore hard choices between fairness metrics must be made before the technical work of detecting and mitigating unfairness can proceed. A further underlying tension is that fairness may also imply that similar people should be treated similarly, but this is often in tension with the ideal of parity between groups, when base rates for the target variable are different between those groups ~\cite{dwork2012fairness}. Fair machine learning therefore faces an upfront set of conceptual ethical challenges; which measures of fairness are most appropriate in a given context? Which variables are legitimate grounds for differential treatment, and why? Are all instances of disparity between groups objectionable? Should fairness consist of ensuring everyone has an equal probability of obtaining some benefit, or should we aim instead to minimise the harms to the least advantaged? In making such tradeoffs, should the decision-maker consider only the harms and benefits imposed within the decision-making context, or also those faced by decision-subjects in other contexts? What relevance might past, future or inter-generational injustices have?

Such questions of discrimination and fairness have long been of significant interest to moral and political philosophers, who have expended significant efforts in formalising, differentiating and debating many of the central concepts involved. It is therefore unsurprising that attempts to formalise fairness in machine learning contain echoes of these old philosophical debates. Indeed, some seminal work in the FAT-ML community explicitly refers to political philosophy as inspiration, albeit in a limited and somewhat ad-hoc way.\footnote{Notable examples include references to the work of authors such as H. Peyton Young, John Rawls, and John Roemer in ~\cite{dwork2012fairness,joseph2016rawlsian}.} A more comprehensive overview could provide a wealth of argumentation which may usefully apply to the questions arising in the pursuit of more ethical algorithmic decision-making. This article aims to provide an overview of some of the relevant philosophical literature on discrimination, fairness and egalitarianism in order to clarify and situate the emerging debate within the discrimination-aware and fair machine learning literature. Throughout, I aim to address the conceptual distinctions drawn between terms frequently used in the fair ML literature--including `discrimination' and `fairness'--and the use of related terms in the philosophical literature. The purpose of this is not merely to consider similarities and differences between the two discourses, but to map the terrain of the philosophical debate and locate those points which provide helpful clarification for future research on algorithmic fairness, or otherwise raise relevant problems which have yet to be considered in that research.

I begin by discussing philosophical accounts of what discrimination is and what makes it wrong, if and when it is wrong. I show how on certain accounts of what makes discrimination wrong, the proposed conditions are unlikely to obtain in algorithmic decision-making contexts. If correct, these accounts do not necessarily imply that algorithmic decision-making is always morally benign--only that its potential wrongness is not to be found in the notion of discrimination as it is traditionally understood. This leads us to consider other grounds on which algorithmic decision-making might be problematic, which are primarily captured by a variety of considerations connected to the ideals of egalitarianism--the notion that human beings are in some fundamental sense equal and that efforts should be made to avoid and correct certain forms of inequality. This discussion suggests that `fairness' as used in the fair machine learning community is best understood as a placeholder term for a variety of normative egalitarian considerations. Notably, while egalitarianism is a widely held principle, exactly what it requires is the subject of much debate. I provide an overview of some of this debate and finish with implications for the incorporation of `fairness' into algorithmic decision-making systems.

\section{What is discrimination, and what makes it wrong?}

Early work which explored how to embed fairness constraints in machine learning used the term `discrimination-aware' rather than `fair'. While this terminological difference may seem relatively insignificant amongst computer scientists, it points to a potentially important distinction which has been the preoccupation of much philosophical writing on the topic. For philosophers, the division of moral phenomena into conceptually coherent categories is both intrinsically satisfying, and, hopefully, brings helpful clarification to otherwise perplexing issues. Getting a closer conceptual grip over what discrimination consists in, what (if anything) makes it distinctively wrong, and how it relates to other moral concepts such as fairness, should therefore help clarify whether and when algorithmic systems can be wrongfully discriminatory and what ought to be done about them. It may turn out that so-called `algorithmic discrimination' differs in important ways to classical forms of discrimination, such that different counter-measures are appropriate.

\subsection{Mental state accounts}

Paradigm cases of discrimination include differential treatment on the basis of membership of a salient social group--e.g. gender or `race'--by those with decision-making power to distribute harms or benefits. Examples include employers who prefer to hire a male job applicant over an equally qualified female applicant, or parole officers who impose stricter conditions on parolees of a particular minority in comparison to a privileged majority. Focusing on these paradigm cases, a range of accounts of what makes discrimination wrong place emphasis on the intentions, beliefs and values of the decision-maker. For such mental state accounts, defended by among others Richard Arneson, Thomas Scanlon, and Annabelle Lever, the existence of systematic animosity or preferences for or against certain salient social groups on the part of the decision-maker is what makes discrimination wrong ~\cite{arneson1989equality,lever2016racial,scanlon2009moral}. Such concerns might be couched in terms of the defective moral character of the decision maker--that they show bad intent or animosity, for example-or in terms of the harmful effect of such intentions on the discriminated-against individual, such as humiliation as a result of lack of respect.

On such accounts, the decision-maker's intent is key to discrimination. A decision-maker with no such intent, who nonetheless accidentally and unwittingly created such disparities, would arguably not be guilty of wrongful discrimination (even if the situation is morally problematic on other grounds). Such cases--often called indirect or institutional discrimination in the UK, or in the U.S., disparate impact--might still count as discriminatory on a mental state account of discrimination if the failure of decision-makers to anticipate such disparities, or to redress them when they become apparent, is sufficiently similarly morally objectionable to the failure to treat people equally in the first instance. However, if such conditions do not obtain, then indirect discrimination, while it may be wrong, may not usefully be described as an instance of discrimination at all ~\cite{eidelson2015discrimination}.

This line of thinking suggests a potential challenge to the notion of algorithmic discrimination. If the possession of certain mental states by decision-makers is a necessary condition of a decision being discriminatory, one might argue that algorithmic decision-making systems can never be discriminatory as such, because such systems are incapable of possessing the relevant mental states. This is not the place for discussion of the possibility of machine consciousness, but assuming that it is not (yet) a reality, it seems that AI and autonomous decision-making systems cannot be the bearers of states such as contempt, animosity, or disrespect that would be required for a decision to qualify as discriminatory on such accounts.

That said, proponents of mental state accounts might similarly argue that algorithmic decision-making systems might involve discrimination, without imputing the algorithm with mental states. First, they might argue that human decision-makers, and data scientists responsible for building decision models, might sometimes be condemned if they possess bad intentions which result in them intentionally embedding biases in their system, or if they negligently ignore or overlook unintended disparities resulting from it. Second, as social epistemologists have argued, we can sometimes still morally evaluate decisions which do not stem from one individual but are the result of multiple individual judgements aggregated in variously complex ways ~\cite{gilbert2004collective,pettit2007responsibility}. Drawing from work on judgement aggregation problems in economics and social choice theory, they argue that when suitably arranged in institutional decision-making scenarios, groups of people can be held morally responsible for their collective judgements. Machine learning models trained on data consisting of previous human judgements might therefore be critiqued on similar grounds if those individual judgements were themselves the result of similarly discriminatory intent on the part of those individuals.

However, aside from such special cases, it seems that mental state accounts of discrimination do not naturally transfer to the context of algorithmic decision-making. If these accounts are correct, we might therefore conclude that algorithmic decision-making, while potentially morally problematic, should not be characterised as an example of wrongful discrimination. Alternatively, other accounts of why discrimination is (sometimes) wrongful which are not based on mental states of the decision-maker, might be better able to accommodate discrimination of an algorithmic variety.

\subsection{Failing to treat people as individuals}

One such account, which has received significant attention in recent writing on discrimination, locates the wrongness of discrimination--in both its direct and indirect varieties--in the extent to which it relies on inferences about individuals based on generalisations about groups of which they are a member ~\cite{lippert2014born}. This objection is frequently raised in response to what is often called statistical discrimination ~\cite{phelps1972statistical}. Statistical discrimination is the use of statistical generalisations about groups to infer attributes or future behaviours of members of those groups. For instance, an employer might read evidence purporting to show that smokers are generally less hard-working than non-smokers, and reject a job application from a smoker and give the job a less qualified non-smoker who is anticipated to be more productive.

As economists have argued on the basis of models, such generalisations about groups can be an efficient means for firms to reduce risk when more direct evidence about an individual is lacking ~\cite{phelps1972statistical}. Despite the potential efficiency benefits of generalisation, it is widely regarded as wrong in at least some cases. While intuitions about the wrongfulness of statistical discrimination are widely shared, it has proven surprisingly difficult to articulate coherent objections to the practice, particularly when we go beyond simple cases in which the inference is simply unsupported by rigorous statistical analysis, to those where membership of a group really does correlate with an outcome of interest to the decision-maker. Since algorithmic decision-making could be regarded as a form of generalisation on steroids, any account of discrimination which is grounded in the wrongness of generalisation would be particularly pertinent to our present concerns.

On one popular account, statistical discrimination is wrong, even if the generalisations involved have some truth, because it fails to treat the decision-subject as an individual. Subjecting travellers from Muslim-majority countries to harsher border checks on the basis of generalisations (whether true or false) about the preponderance of terrorism amongst such groups fails to consider each individual traveler on their own merit. Similarly, rejecting a job applicant who smokes because of evidence that smokers are on average (let us assume, for the sake of argument) less productive, unfairly punishes them as a result of the behaviour of other members of that group. Such examples have led some to ground objections to statistical discrimination in its failure to treat people as individuals. If this is correct, it presents a strong challenge to the very existence of algorithmic decision-making systems; since they fail to treat people as individuals by design. Given any two individuals with the same attributes, a deterministic model will give the same output, and would therefore, on this account, be quintessentially discriminatory.

However, others have argued that the failure to treat people as individuals cannot plausibly be the essence of wrongful discrimination ~\cite{schauer2009profiles,dworkin1981equality,lippert2014born}. One concern is that this criterion is too broad because it encompasses generalisation against any kind of group, not just those categories enshrined in human rights law such as gender, `race', religion, etc. While such categories readily trigger concerns about wrongful discrimination, other categories like `smoker' do not obviously invoke discrimination concerns. This suggests that it is only generalisations about certain groups that matter vis-à-vis discrimination. Others object that the very notion of `treating someone as an individual' is misconceived; they argue that, on closer inspection, even decisions which appear to consider the individual are in fact a disguised form of generalisation ~\cite{schauer2009profiles}. Suppose that rather than using `smoker / non-smoker' as a predictor of productivity, the employer requires applicants to undergo some test which more accurately predicts productivity; even in this case, as Schauer argues, the employer must rely upon generalizations from test scores to on-the-job behavior. Test scores might be a more accurate predictor than `smoker / non-smoker', but they are still fundamentally a form of generalization (~\cite{schauer2009profiles}, p68). Unless the test is perfect, some people who perform badly on it would nevertheless turn out to be relatively productive on-the-job.

If this line of critique is correct, then it cannot be the case that treating people differently on the basis of generalisations about a category they fit into is necessarily discriminatory in any wrongful sense. What appear to be criticisms of generalization in general(!), may in fact boil down to criticisms of ~\emph{insufficiently precise} means of generalization. If the border security system (or the recruitment process) could identify more accurate predictor variables, which resulted in fewer burdens on innocent tourists (or hard-working smokers), then the charge of discrimination might lose some force. Of course, more accurate predictions are likely to be more costly, and as such tradeoffs must be made between the harms and benefits of generalization; but either way, this view suggests that anti-discrimination does not necessarily require treating people as individuals. Rather, statistical discrimination may be more or less morally permissible depending on who and how many people are wrongly judged on the basis of membership of whatever statistical groups they may be part of, compared to the costs involved in improving the accuracy of the generalisation. If correct, this should be a welcome conclusion for proponents of algorithmic systems, since they are essentially based on generalisations in some form or other.

So far I have presented arguments which suggest that there are difficulties with accounting for algorithmic discrimination in terms of the wrongness of mental states or of generalisations. Some of these difficulties are internal to the philosophical accounts of discrimination, and others stem from the dis-analogy between human and algorithmic decision-makers. If the wrongness of (algorithmic) discrimination does not consist in the morally suspect intentions of decision-makers, or in failing to treat people as individuals, then what might it consist in? A more general set of egalitarian norms might provide a better footing for a theory of algorithmic fairness. \footnote{However, even if philosophical accounts of discrimination do not easily apply to algorithmic decisions, calling an algorithmic system  discriminatory, (or specifically sexist, racist, etc.) might still be justified by its rhetorical power, or as useful shorthand in everyday discourse.}

\section{Egalitarianism}

Broadly speaking, egalitarianism is the idea that people should be treated equally, and (sometimes) that certain valuable things should be equally distributed. It might seem entirely obvious that what makes discrimination wrongful is something to do with egalitarianism. However, this connection has, perhaps surprisingly, been resisted by many of the previously mentioned theorists of discrimination, with one even claiming that any `connection between anti-discrimination laws and equality it is at best negligible, and in any event is insufficient to count as a justification' ~\cite{holmes2005anti}. Meanwhile, others argue the opposite: that only a direct appeal to egalitarian norms can satisfactorily account for everything that is wrong about discrimination ~\cite{segall2012s}. For our current purposes, this debate can be safely sidestepped. This is not because it is uninteresting or unimportant as a philosophical project. Rather, our purpose here is to examine how egalitarian norms might provide an account of why and when algorithmic systems can be considered unfair; whether or not such unfairness should rightfully be considered a form of discrimination, per se, is not our concern.

This section therefore provides a brief overview of some major debates within egalitarianism and draws out their significance for fair machine learning. 

\subsection{The currency of egalitarianism and spheres of justice}

Invariably in machine learning contexts where fairness issues arise, a system is mapping individuals to different outcome classes which are assumed to have negative and positive effects; such as being approved / denied a financial loan, high or low insurance prices, or a greater or fewer number of years spent under incarceration. We assume that these outcome classes are means or barriers to some fundamentally valuable object or set of objects which ought to be to some extent equally distributed. But what exactly is the `currency' of egalitarianism that lies behind the valuation of these outcome classes? Egalitarians have articulated various competing views, including welfare, understood in terms of pleasure or preference-satisfaction ~\cite{cohen1989currency}; resources such as income and assets ~\cite{rawls2009theory,dworkin1981equality}; or capabilities, understood as both the ability and resources necessary to do certain things ~\cite{sen1992inequality}. Others propose that inequalities in welfare, resources, or capabilities may be acceptable, so long as citizens have equal political and democratic status ~\cite{anderson1999point}.

The question of what egalitarianism is concerned with (the `equality of what?' debate as it is sometimes referred to), is relevant to our assumptions about the impact of different algorithmic decision outcomes. In many cases -- like the automated allocation of loans, or the setting of insurance premiums -- the decision outcomes primarily affect distribution of resources. In others -- like algorithmic blocking or banning of users from an online discussion -- the decisions may be more directly related to distribution of welfare or capabilities. The importance of each currency of egalitarianism may plausibly differ between contexts, thus affecting how we account for the potentially differential impacts of algorithmic decision-making systems.

This debate also trades heavily on the intuition that different people may value the same outcome or set of harms and benefits differently; yet most existing work on fair machine learning assumes a uniform valuation of decision outcomes across different populations. In some cases it may be safe to assume that different sub-groups are equally likely to regard a particular outcome class -- e.g. being denied a loan, or being hired for a job - as bad or good. But in other cases, especially in personalisation and recommender systems where there are multiple outcome classes with no obvious universally agreed-on rank order, this assumption may be flawed.

A connected debate concerns whether we should apply a single egalitarian calculus across different social contexts, or whether there are internal `spheres of justice' in which different incommensurable logics of fairness might apply, and between which redistributions might not be appropriate ~\cite{walzer2008spheres}. For instance, with regard to civil and democratic rights like voting in elections, the aim of egalitarianism might be absolute equal distribution of the good, rather than merely equality of opportunity to compete for it. This idea would explain the intuition that voter registration tests are wrong, while tests for jobs are not. Requiring some form of test prior to voting may ensure equality of opportunity in the sense that everyone has an equal opportunity to take the test; but since talent and efforts are not equally distributed, some people may fail the test, and there would not be equality of outcome. But an essential element of democracy, one might argue, is that voting rights shouldn't depend on talent or effort. However, when it comes to competition for social positions and economic goods, we may be concerned with ensuring equality of opportunity but less concerned about equality of outcome. We may consider it fair, other things being equal, that the most qualified applicant obtains the role, and that the most industrious and / or talented individuals deserve more economic benefits than others (even if we believe that current systems do not actually ensure a level playing field, and some level of income redistribution is also morally required).

Different contexts being subject to different spheres of justice would have a direct bearing on the appropriateness of certain fairness-correcting methods in those contexts. Equality of opportunity may be an appropriate metric to apply to models that will be deployed in the sphere of `economic justice'; e.g. selection of candidates for job interviews or calculation of insurance. But in contexts which fall under the sphere of civil justice, we may want to impose more blunt metrics like equality of outcome (or `parity of impact'). This might be the case for airport security checks, where it is important to a sense of social solidarity that no group is over-examined as a result of a predictive system, even if there really are differences in base rates ~\cite{hellman2008discrimination}. We therefore can't assume that fairness metrics which are appropriate in one context will be appropriate in another.

\subsection{Luck and desert}

A second major strand of debate in egalitarian thinking considers the role of notions like choice ~\cite{huseby2016can} and desert ~\cite{temkin1986inequality} in determining which inequalities are acceptable? In which circumstances, and to what extent, should people be held responsible for the unequal status they find themselves in? So-called `luck egalitarians' aim to answer this question by proposing that the ideal solution should allow those inequalities resulting from people's free choices and informed risk-taking, but disregard those which are the result of brute luck ~\cite{arneson1989equality}. As free-willed individuals, we are capable of making choices and bearing their consequences, which may sometimes make us better or worse off than others. The choices we make may deserve certain rewards and punishments. However, where inequalities are the result of circumstances outside an individual's control (e.g. being born with a debilitating health condition, or being born into a culture in which one's skin colour results in systemically worse treatment), egalitarians argue for their correction. However, defining a principle which can demarcate between those inequalities which are and are not chosen or deserved is a tricky prospect, one which has vexed egalitarians for centuries.

The luck egalitarian aim of pursuing redistribution only where inequalities are due to pure luck, and leaving in place inequalities which are the result of personal choices and informed gambles, raises interesting questions for which variables should be included in fair ML models. High-profile controversies around the creation of criminal recidivism risk scoring in the US, notably the COMPAS system, have focused primarily on the differential impacts on African American and Caucasian subjects ~\cite{angwin2016machine}. But one of the potentially objectionable features of the COMPAS scoring system was not the use of `race' as a variable (which it did not), but rather its use of variables which are not the result of individuals' choices, such as being part of a family, social circle, or neighbourhood with higher rates of crime. These may be objectionable in part because they correlate with `race' in the U.S., but they are also objectionable more generally to the extent that they are not the result of personal choices. As such, any inequalities that arise from them should not, on the luck egalitarian view, be tolerated.

Furthermore, as critics of luck egalitarianism have argued, sometimes even inequalities which are the result of choice still ought to be compensated. For instance, as Elizabeth Anderson has argued, standard luck egalitarianism leads to the vulnerability of dependent caretakers, because it would not compensate those who are responsible for choosing to care for dependents rather than working a wage-earning job full time ~\cite{anderson1999point}. This is what Thayson and Albertson call a `costly rescue' ~\cite{thaysen2017bad}; on their view, luck egalitarianism should only be sensitive to responsibility for creating advantages and disadvantages -- not to responsibility for distributing them. Thus, individuals who voluntarily place themselves in unequal positions might sometimes deserve compensation if their choice served some important social purpose. To return to the COMPAS case, even if certain variables like one's social circle or neighbourhood were the result of individual choice (which is perhaps more likely to be the case for the economically advantaged), such choices may nevertheless deserve protection from negative consequences. This might apply in cases like those outlined by Anderson, Thayson and Albertson above; for instance, one might choose to remain a resident in a high crime neighbourhood in order to make a positive difference to the community.

\subsection{Deontic justice}

In applying egalitarian political philosophy to the analysis of particular instances of inequality, these abstract principles need to be supplemented with empirical claims about how and why certain circumstances obtain. This reflects the sense in which egalitarianism can be, in Derek Parfit's terms, deontic; that is, not concerned with an unequal state of affairs per se, but rather with the way in which that state of affairs was produced ~\cite{parfit1997equality}. Where analytic philosophy ends, sociology, history, economics, and other nascent disciplines are needed, to understand the specific ways in which some groups come to be unfairly disadvantaged ~\cite{fang2010theories,hooks1992yearning}. Only then can we meaningfully evaluate whether and to what extent a given disparity is unfair. But consideration of historical and sociological contexts can also inform philosophical accounts and raise new questions. One example is in the attribution of responsibility; who should be held responsible for the initial creation of inequalities, and who should be held responsible for correcting them? Can historical injustices perpetrated by an institution in the past ground present-day claims for redistribution? Is a particular instance of unequal treatment of a particular group worse if it takes place in a wider context of inequality for that group, compared to a general pattern of benign or advantageous treatment?

Abstracting away from particular cases and considering broader historical and social trends may better enable us to account for what makes certain forms of unequal treatment particularly concerning. In discussing what makes racial profiling worse than other forms of profiling, even if it is based on statistical facts, Kasper Lippert-Rasmussen argues (~\cite{lippert2014born}, p. 300):
\begin{quote}
`Statistical facts are often facts about how we choose to act. Since we can morally evaluate how we choose to act, we cannot simply take statistical facts for granted when justifying policies: we need to ask the prior question of whether it can be justified that we make these statistical facts obtain'
\end{quote}
On this view, the particular wrongness of racial profiling can only be understood by appealing to the social processes which cause the statistical regularities to obtain in the first place. In Lippert-Rasmussen's example, the reason racial profiling for crime in the U.S. `works' (if it does at all), may be due to things that the white majority do or fail to do which might reduce or eliminate the reliability of such statistical inferences.

Similarly, such `deontic', historical and sociological considerations can provide critical background information which is likely to be crucially relevant in determinations of fairness in particular algorithmic decision-making contexts. Historical causes of inequalities and broader existing social structures cannot be ignored when deploying models in such contexts. At a basic level, this means that feature selection--both for protected characteristics and other variables--should be informed by such information, but it also might determine which disparities take priority in fairness analysis and mitigation if more than one exists. More broadly, deontic considerations may help situate and illuminate the moral tensions that arise between different and incompatible fairness metrics. Returning to the debate about the COMPAS criminal recidivism scoring system, where unequal base rates mean that accuracy equity is mathematically impossible to achieve alongside equalised false positive rates, a deontic egalitarian perspective suggests focusing attention on the historic reasons for such unequal base rates. While this may not in itself directly resolve the question of which fairness metric ought to apply, it does suggest that part of the answer should involve consideration of the historic and contemporary factors responsible for the broader social situation from which the incompatibility dilemma arises.

\subsection{Distributive versus representative harms}

Finally, it is important to note that some aspects of egalitarian fairness may not be distributive in a direct way, in the sense that they concern the distribution of benefits and harms to specific people from specific decisions. Rather, they may be about the fair representation of different identities, cultures, ethnicities, languages, or other social categories. For instance, states with multiple official languages may have a duty to ensure equal representation of each language, a duty which need not derive from any specific claims about the unequal benefits and harms to individual members of each linguistic group ~\cite{taylor1994multiculturalism}. Similar arguments might be made about cultural representation in official documents, or even within the editorial policies of institutions in receipt of public money. Even private institutions might well voluntarily impose such duties upon themselves. There is a debate about the extent to which equal cultural recognition and distributive egalitarianism are truly distinct notions; some argue that `recognition and distribution are two irreducible elements of a satisfactory theory of justice', while others that `any dispute regarding redistribution of wealth or resources is reducible to a claim over the social valorisation of specific group or individual traits' (~\cite{mcqueen2011social} in discussion of ~\cite{fraser2003redistribution}).\footnote{An earlier version erroneously attributed these quotes to \cite{fraser2003redistribution}. I thank Dr. Joel Anderson for bringing this to my attention.}

Such notions of representational fairness capture many of the most high-profile controversial examples of algorithmic bias. For instance, much-reported work on gender and other biases in the language corpora used to produce word embeddings is an example of representational fairness ~\cite{bolukbasi2016man,caliskan2016semantics}. In such cases, the problem is not necessarily one of specific harms to specific members of a social group, but rather one of the way in which certain groups are represented in digital cultural artefacts, such as natural language classifiers or search engine results. This may require different ways of approaching fairness and bias since the notions of differential group impacts and treating like people alike do not apply; instead, the goal might be to ensure equal representation of groups in a ranking ~\cite{zehlike2017fa}, or to give due weight to different normative / ideological outlooks in a classifier which automates the enforcement of norms ~\cite{binns2017like}.

\section{Conclusion}

Current approaches to fair machine learning are typically focused on interventions at the data preparation, model-learning or post-processing stages. This is understandable given the typical remit of data scientists who are intended to carry out these processes. However, there is a danger that this results in an approach which focuses on a narrow, static set of prescribed protected classes, derived from law and devoid of context, without considering why those classes are protected and how they relate to the particular justice aspects of the application in question. Philosophical accounts of discrimination and fairness prompt reflection on these more fundamental questions, and suggest avenues for further consideration of what might be relevant and why. 

This raises a series of practical challenges which may limit how effective and systematic fair ML approaches can be in practice. Attempting to translate and elucidate the differences between such egalitarian theories in the context of particular machine learning tasks will likely be tricky. In simple cases, it may be that feature vectors used to train models include personal characteristics which can intuitively be classed as either chosen or unchosen (and therefore legitimate or illegitimate grounds for differential treatment according to e.g. luck egalitarianism). But more often, a contextually appropriate approach to fairness which truly captures the essence of the relevant philosophical points may hinge on factors which are not typically present in the data available in situ. Such missing data may include the protected characteristics of affected individuals ~\cite{veale2017fairer}, but also information relevant to an assessment of an individual's responsibility, culpability or desert--such as their socio-economic circumstances, life experience, personal development, and the relationships between them. Attempts to draw such conclusions from training data and lists of legally protected categories alone, are unlikely to do justice to the way that questions of justice arise in idiosyncratic lives and differing social contexts.

\acks{The author was supported by funding from the UK Engineering and Physical Sciences Research Council (EPSRC) under \emph{SOCIAM: The Theory and Practice of Social Machines}, grant number EP/J017728/2, and \emph{PETRAS: Cyber Security of the Internet of Things}, under grant number EP/N02334X/1.}

\bibliography{jmlr-sample}

\appendix





\end{document}